%% file: paper5.tex
\documentstyle[aps,epsf,array]{revtex}
\newcommand{\newsection}[1]{
\vspace{2mm}
\pagebreak[3]
\addtocounter{section}{1}
\setcounter{subsection}{0}
\setcounter{footnote}{0}
\large {\bf\thesection. #1 \\}
\normalsize
\nopagebreak
\medskip
\nopagebreak
\hspace{3mm}}
\newcommand{\be}{
\begin{equation}}
\newcommand{\ee}{
\end{equation}}
\newcommand{\bfig}{
\begin{figure}}
\newcommand{\efig}{
\end{figure}}
\newcommand{\bcen}{
\begin{center}}
\newcommand{\ecen}{
\end{center}}
\newcommand{\bea}{
\begin{eqnarray}}
\newcommand{\eea}
{\end{eqnarray}}

\begin{document}
\addtolength{\baselineskip}{.7mm}
\thispagestyle{empty}
\begin{center}
{\Large{\bf  Highly Designable Protein Structures  \\
and \\
Inter Monomer Interactions }} \\[5mm]
{\bf M. R. Ejtehadi$^{1,2}$\footnote{
\it e-mail: reza@netware2.ipm.ac.ir},}
{\bf N. Hamedani$^1$, H. Seyed-Allaei$^1$,} \\
{\bf V. Shahrezaei$^1$} and {\bf M. Yahyanejad$^1$} \\

$^1${\it Department of Physics, Sharif University of Technology,}\\
{\it Tehran P. O. Box: 11365-9161, Iran.}\\
and \\
$^2${\it Institute for Studies in Theoretical Physics and Mathematics}\\
{\it  Tehran P. O. Box: 19395-5531, Iran.}\\
\end{center}
%\pagebreak
%
%{\parbox{14cm}{\hspace{5mm}
%\begin{center}
%{\bf Abstract}\\[5mm]
%\end{center}
%\small
\begin{abstract}
\\
 By  exact computer enumeration and combinatorial methods, we have
 calculated the
designability of proteins
in  a simple lattice  {\bf H-P} model for the protein folding problem.
We show that if the strength of the
non-additive part of the interaction potential becomes
larger than a critical value, the degree of designability
of structures will depend on  the
parameters of potential.
 We also show that the existence of a unique ground state is
 highly  sensitive to mutation in certain sites.
\end{abstract}
\nopagebreak
\vfill
\newpage
\setcounter{section}{0}
\setcounter{equation}{0}
%%\numberbysection
%
%%%%%%%%%%%%%%%%%%%%%%%%%%%%%%%%%%%%%%%%%%%%%%%%%%%%%%%%%%%%%%%%%%%
%
%%%%%%%  Section 1  %%%%%%%%%%%%%%%%%%%%%%%%%%%%%%%%%%%%%%%%
%
\newsection{Introduction}

Biologically active proteins fold into a native
 compact structure despite the
huge number of possible configurations \cite{1}.
Though the mechanism of protein folding is not fully
understood, it has been known since the  re-folding
experiments of Anfinsen \cite{Anfinsen1} that globular proteins fold in the absence
of any catalytic biomolecules. From this fact,
 it has been established  that for proteins,the three
dimensional folded  structure  is the minimum free energy
structure, and,
the information coded
in the amino-acid sequence is sufficient to determine
the native structure \cite{Anfinsen2}.
 The compactness of this unique native state is largely due
 to the existence of an optimal amount of hydrophobic amino-acid
 residues \cite{4}, since these biological objects are usually designed to
 work in water \cite{Garel}.
The relation  between the primary one dimensional sequence and the final
compact three dimensional structure is the task of the
 protein folding problem.

In addition to the paradoxical problem of kinetics and time scales of the
folding process \cite{Levinthal}, there is another mystery.
If proteins are made randomly by amino acids,
the number of all possible  such  proteins with typical length  of
$100$, is far larger than the number of proteins which  actually
 occur in nature. One
hypothesis is that the naturally selected sequences are special because
they are coded for structures that have unique and stable native states,
allowing for easy folding.
Thus a 
 central question of protein evolution is how mutational
change in the amino acid sequence leads to changes in the structure
 and stability.

Some efforts have been made in order to  study 
the stability of proteins against
 mutation by searching the 
 two dimensional configuration space \cite{6,7}.
One simple model  used in  these studies is the {\bf H-P} model \cite{8}.
In this model there are only two types of chain monomers,
 hydrophobic ({\bf H}) and
polar ({\bf P}). Every {\bf H-H} contact between topological neighbours is
assigned a   negative contact energy, and other contact interactions 
are set to zero.

Recently  Li {\it et al.} \cite{Li}, have looked at this problem in three
dimensions.
Calculating the  energy of all possible 27-mers in all compact
three dimensional configuration, they have found that, there are a few
structures, into  which a high number of sequences uniquely fold.
This structures were named  "highly designable"  and the
number of sequences which fold into each state  was 
named its  "designability".
In their {\bf H-P} model, they choose the contact energy between
{\bf H} and {\bf P} monomers by some physical arguments \cite{Li,10}.
Other  significant points of their work are:
a) Only a few percent of sequences have unique ground state;
b) There is a jump in energy gap for these highly designable structures.
Thus the highly designable structures are more stable against
mutation and thermal fluctuation.

Dill and Chan \cite{11} have argued that many of the phenomena observed
in proteins can be adequately understood in terms of the H-P model,
but according to the
 work of Pande {\it et al.} \cite{12} the designability of a
 conformation does depend
on the nature of interactions between monomers. May be any
 interaction leads to some highly designable structures,
 but different interactions yield different patterns.

In our work we study this problem  
for an additive potential. We will show that there
are some highly designable structures for this potential,
but the low designable structures will   disappear because of degeneracy of 
ground state. We will show that there is a ladder structure for energy 
levels for this form of potential.
We then add a non-additive part to the energy,
then the ground state degeneracy of  low designable structures
will be removed. 
We  show there is a critical  value for non-additive
 part of potential, where
below this critical value
the patterns of highly designable structures are fixed,
but  above  this critical value the 
 designability of structures is  sensitive to the
value of non-additive part of the potential.
We  show that the sequences which fold to highly
designable structure are sensitive to mutation of some  sites.

An  additive  potential has the following advantages:

a) It allows us to prepare a very fast algorithm which is 
then possible to run on a PC.

b) It enables us to solve and study some parts of the 
problem by combinatorial methods.

c) It gives a clear picture for designability.

d) A ladder spectrum for the energy levels results, thus  it arms us to
study the problem for non-additive potentials.

\newpage
%
%%%%%%%%%%%%%%%%%%%%%%%%%%%%%%%%%%%%%%%%%%%%%%%%%%%%%%%%%%%%%%%%%%%
%
%%%%%%%  Section 2  %%%%%%%%%%%%%%%%%%%%%%%%%%%%%%%%%%%%%%%%
%
\newsection{ The Model}

We consider an {\bf H-P}  lattice model \cite{8}.
In this model only non-sequential nearest
neighbours interact. Because the native structures of proteins are compact with
the {\bf H} type monomers  sitting in the core,
the effective potentials which are usually  used,
all of the forces are attractive (negative values 
for potential) and the strength
of the force between {\bf H-H} monomers is greater than others.
We can write the general form of the potential in an arbitrary
energy scale as:

\be
E_{\bf PP}=0, \hspace{10mm} E_{\bf HP}=-1, \hspace{10mm} E_{\bf HH}=-2-\gamma.
\ee

The most usual choice of {\bf H-P} model potential corresponds to the 
 limit $\gamma\gg 1$
\cite{6,7,8,11}, however physical arguments are consistent with
a smaller value for $\gamma$, for instance $\gamma=0.3$  
was used by Li {\it et al.}  \cite{Li}. 
They have calculated the energy of all of $2^{27}$ sequences 
in $103,346$ compact configuration for a $27$-sites cube, by a 
huge enumeration.

In the case $\gamma=0$, we have an additive potential.
If we let  ${\bf H}=-1$, and ${\bf P}=0$, 
we can rewrite the potential  
in the form,
\be 
E_{\sigma_i \sigma_j}=\sigma_i + \sigma_j .
\ee

Following Li {\it et al.} \cite{Li}, 
we consider only compact structures of sequences with length 27,
occupying all sites of a $3\times3\times3$ cube \cite{shakh}.
There are $103,346$ compact configurations which are not related to
each other by rotation and reflection symmetries.
Let us call the set of all compact  structures, the structure space.

A protein of length
$N$ may be shown by an $N$-component vector
\be
|\sigma\rangle = |\sigma_{i_1}, \sigma_{i_2}, \dots ,\sigma_{i_N}\rangle ,
\ee
where $i_n=1,2$ refers to {\bf P} and {\bf H} residues.
Thus the number of  such $N$-component vectors
for  proteins with length $27$ 
is $2^{27}$. 
Let us call the set of  $|\sigma\rangle$,  the sequence space.

Because of the  additive form of the potential, we can write the energy of
a given $|\sigma\rangle$ in any spatial configuration as,
\be
E=\sum_{i=1}^{27} p_i \sigma_i,
\ee
where $p_i$'s are the number of non-sequential neighbours of
the $i$th monomer, or by introducing the neighbourhood vector $|P\rangle$,
\be
E=\langle\sigma|P\rangle.
\ee

The vector $|P\rangle$ has 27 components and at $i$th
component has the number of neighbours of the $i$th monomer.
Due to the shape of $|P\rangle$ the type of neighbours is not
relevant and all we have to do is count the non-sequential neighbours.
This gives us an additional symmetry for the energy  
that is different from spatial symmetries.
For example any of the sites in a two dimensional $5\times5$
square for two spatial configurations which are shown in figs. 1a and 1b,
have equal neighbours, but the labels of their neighbours
 are not the same.
Visualisation of the same effect in 3 dimensions is a
bit harder, but it dose exist.

The space of all $3$ dimensional structures 
 has $103,347$ members for all compact full filled structures in 
a $3\times 3\times 3$ cube. Due to this additional symmetry this space  is
divided into $6291$ subspaces, where all members of each subspace have the
same $|P\rangle$. Let the number of members of a subspace be, 
$N_d$. The range of $N_d$ is from $1$ to 
$96$.
Fig. 2, shows that the frequency  of large values of $N_d$, is low.
Interestingly there are a lot of $|P\rangle$'s
which only point to one structure.
\bfig
\bcen
\input{fig1.tex}
\ecen
\caption{
		The number of neighbours for corresponding sites 
		in these two configuration are the same, but the 
                neighbours are not, for instance, site 18 in (a)
		is the neighbour of 5, but they are not adjacent 
		in (b).}
\efig

\bfig
\bcen
\epsfxsize=10.0cm
\epsfbox{fig2.eps}
\ecen
\caption{
		Histogram of $N_d$ for members of structure 
                space. It is interesting that there are some $P$ sets
		with $N_d=1$.}

\efig
We have  calculated the   energy of  all $2^{27}$ $| \sigma\rangle$ on all
$|P\rangle$. We find the degeneracy of ground state in 
structure space. One can see the distribution of number
 of ground state degeneracies $g$,
for all of $2^{27}$ sequences in fig. 3.
There are only a few  sequences which 
have non-degenerate ground state,
this corresponds to the $8.47\%$ of sequences at $g=1$.
If energy of
one sequence is minimised in a $|P\rangle$ with $N_d$ greater  than one
it has degenerate ground state.
According to definition of designability, such sequences should not
be considered.  
The distribution of $N_s$ is presented for $\gamma=0$ in fig. 4.
Comparing this figure with fig. 2 of Li {\it et al.} \cite{Li},
we observe that there is no similarity.
This suggests that designability ($N_s$), is sensitive to the
value of $\gamma$, which is $\gamma=0$ in our work, whereas
Li {\it et al.} chose $\gamma= 0.3$.
However as we shall see later, the fact that at $\gamma=0$,
we have an additive potential plays an important role. In fact a small
value of $\gamma$ radically change the picture.

\bfig
\bcen
\epsfxsize=10.0cm
\epsfbox{fig3.eps}
\ecen
\caption{
		Histogram of degeneracy of ground state.
		The sequences which have non-degenerate
		ground state, correspond only to $g=1$, 
		in this diagram.}

\efig
\vspace{4mm}
\bfig
\bcen
\epsfxsize=10.0cm
\epsfbox{fig4.eps}
\ecen
\caption{
		Histogram of $N_s$ for additive potential.}

\efig

If we  consider all of sequences which have non-degenerate ground state
in structure space,  we get a new picture for designability.
This means that we calculate the designability of all $|P\rangle$'s,
and not only those with $N_d=1$.
This is contrast to $N_s$ which had only $N_d=1$.

 To recognise this difference,
 we show designability of structures, by $N'_s$.
Fig. 5 shows the distribution of $N'_s$.
Many of points in this fig. 5  are related to some $|P\rangle$'s with
$N_d \neq 1$.  
 We shall use this picture to express the nature of
the energy gap in the case $\gamma \neq 0$
in  section {\bf V}.
\bfig
\bcen
\epsfxsize=10.0cm
\epsfbox{fig5.eps}
\ecen
\caption{
		Histogram of $N_s'$ for additive potential.
		Note that many of points in this diagram 
                correspond to some $|P\rangle$'s which point  
		to more than one spatial configuration.}

\efig

In our enumeration
we have calculated the energy of any sequence in all $6291$ $|P\rangle$'s,
but in fig. 5 , we show the results for $3153$ $|P\rangle$'s which are not
related to each other by reverse labelling.
We can not reduce the structure space  according to this symmetry
before enumeration. Reverse labelling
for a nonsymmetric  sequence gives two
different configurations which  may have different energies.

%%%%%%%%%%%%%%%%%%%%%%%%%%%%%%%%%%%%%%%%%%%%%%%%%%%%%%%%%
\newsection{Energy levels}

The number of non-sequential neighbours is related to type of site.
A $3\times3\times3$ cube has 8 corner sites $(C)$, 12 link sites $(L)$,
6 face sites $(F)$, and one centre site $(O)$ (fig. 6).
$C$ sites have three neighbours, where two of them are
 connected by sequential
links and there is only one non-sequential neighbour. Similarly $L$, $F$ 
 and $O$ sites have 2, 3 and 4 non-sequential neighbours respectively. 
 We must add 1 to these numbers for
two ends of chain.
\bfig
\bcen
\input{fig6.tex}
\ecen
\caption{
	       A $3\times3\times3$ cube has 8 corner sites, 
	       12 link sites, 6 face sites and 1 centre site.}

\efig
This sites are divided in two classes, $\{C,F\}$ and $ \{L,O\}$.
In a self avoiding walk in this cube, we must jump in any step  from
one set to other. The first set has 14 members and the second has
13.
Thus a walk passes through $C$ and $F$ sites in odd steps,
and through $L$ and $O$
sites in even steps.
In other words, the odd components of $|P\rangle$ are 1 or 3,
and even components are 2 or 4,
(except the 1st and 27th components which are
like even components).
 Thus,

\be
|P\rangle= |p_1,\dots ,p_{27}\rangle,  \nonumber
\ee
where,
\be
p_i= \left\{ \begin{array}{llr}
		     1,3   & \ \ \ \ \ {\rm odd} &\ i's; \\
		     2,4   & \ \ \ \ \ {\rm  even}&\ i's. \\
		\end{array}
     \right.
\ee
Therefore the energy for a sequence $\sigma_\alpha$ in a structure $P_\mu$
 is
\bea
E_{\alpha\mu} &=& \langle \sigma_\alpha|P_\mu\rangle \nonumber \\
&=&\sum_{i\in {\rm odd}} (p_{\mu i}-1)\sigma_{\alpha i} +
\sum_{i\in {\rm even}} (p_{\mu i}-2)\sigma_{\alpha i} +
\sum_{i\in {\rm odd}}  \sigma_{\alpha i}+ 2\sum_{i\in {\rm even}} \sigma_{\alpha i}.
\eea 
By introducing the new binary variable $x$ the above can be rewritten as,
\be
E_{\alpha\mu}=\sum_{i=1}^{27} 2 x_{\mu i}\sigma_{\alpha i} +
\sum_{i\in {\rm odd}}  \sigma_{\alpha i}+2\sum_{i\in {\rm even}} \sigma_{\alpha i},
\label{xvec}
\ee
where,
\be
x_i= \left\{ \begin{array}{llr}
		     0   & \ \ \ \ \  p_i=1 \ {\rm or} \  2; \\
		     1   & \ \ \ \ \  p_i=3 \ {\rm or} \  4. \\
		\end{array}
     \right.
\ee
Two last terms in eq. (\ref{xvec}) are independent of $|X\rangle$
or $|P\rangle$, thus they result in a constant,
 which can be ignored  when  comparing  energies of a
sequence in different configurations.
The first term in eq. (\ref{xvec}) is an integer times two,
thus it results in  a ladder
energy spectrum with gaps of 2. Therefore the energy gap for
all of structures is the same, and there is no difference between 
low designable and high designable structures.

%%%%%%%%%%%%%%%%%%%%%%%%%%%%%%%%%%%%%%%%%%%%%%%%%%%%%%%%%%%%
\newsection{Combinatorial Approach}

Our aim is  to find the $N'_s$ for any
spatial configuration, determined by a vector  $|P\rangle$.
Because $|X\rangle$ has a  simpler structure, than  $|P\rangle$,
we shall use
 $|X\rangle$ instead of  $|P\rangle$.
Any vector $|X\rangle$, has seven $1$'s and twenty $0$'s.
One of the $1$'s is in the even sites, and the others
are in odd sites. Energy could be calculated by
performing a ``logical  and'' of two
binary numbers ($|\sigma\rangle$ and $|X\rangle$).
 For example,  a typical
$|P\rangle$ is,

\vspace{5mm}
\unitlength 0.7mm
\linethickness{0.4pt}
\begin{picture}(210.00,10.00)
\put(200.00,5.00){\makebox(0,0)[cc]{$P$.}}
\put(10.00,10.00){\makebox(0,0)[cc]{$4$}}
\put(20.00,10.00){\makebox(0,0)[cc]{$1$}}
\put(30.00,10.00){\makebox(0,0)[cc]{$3$}}
\put(40.00,10.00){\makebox(0,0)[cc]{$3$}}
\put(50.00,10.00){\makebox(0,0)[cc]{$1$}}
\put(60.00,10.00){\makebox(0,0)[cc]{$1$}}
\put(70.00,10.00){\makebox(0,0)[cc]{$1$}}
\put(80.00,10.00){\makebox(0,0)[cc]{$1$}}
\put(90.00,10.00){\makebox(0,0)[cc]{$3$}}
\put(100.00,10.00){\makebox(0,0)[cc]{$1$}}
\put(110.00,10.00){\makebox(0,0)[cc]{$3$}}
\put(120.00,10.00){\makebox(0,0)[cc]{$1$}}
\put(130.00,10.00){\makebox(0,0)[cc]{$3$}}
\put(140.00,10.00){\makebox(0,0)[cc]{$2$}}
\put(15.00,00.00){\makebox(0,0)[cc]{$2$}}
\put(25.00,00.00){\makebox(0,0)[cc]{$2$}}
\put(35.00,00.00){\makebox(0,0)[cc]{$4$}}
\put(45.00,00.00){\makebox(0,0)[cc]{$2$}}
\put(55.00,00.00){\makebox(0,0)[cc]{$2$}}
\put(65.00,00.00){\makebox(0,0)[cc]{$2$}}
\put(75.00,00.00){\makebox(0,0)[cc]{$2$}}
\put(85.00,00.00){\makebox(0,0)[cc]{$2$}}
\put(95.00,00.00){\makebox(0,0)[cc]{$2$}}
\put(105.00,00.00){\makebox(0,0)[cc]{$2$}}
\put(115.00,00.00){\makebox(0,0)[cc]{$2$}}
\put(125.00,00.00){\makebox(0,0)[cc]{$2$}}
\put(135.00,00.00){\makebox(0,0)[cc]{$2$}}
\end{picture}
\\[5mm]
To recognise odd and even components of the vectors, we show them in the
above form, writing the even sites below.
 The vector $|X\rangle$ corresponding to the above $|P\rangle$ is,

\vspace{5mm}
\unitlength 0.7mm
\linethickness{0.4pt}
\begin{picture}(210.00,10.00)
\put(200.00,5.00){\makebox(0,0)[cc]{$X$.}}
\put(10.00,10.00){\makebox(0,0)[cc]{$1$}}
\put(20.00,10.00){\makebox(0,0)[cc]{$0$}}
\put(30.00,10.00){\makebox(0,0)[cc]{$1$}}
\put(40.00,10.00){\makebox(0,0)[cc]{$1$}}
\put(50.00,10.00){\makebox(0,0)[cc]{$0$}}
\put(60.00,10.00){\makebox(0,0)[cc]{$0$}}
\put(70.00,10.00){\makebox(0,0)[cc]{$0$}}
\put(80.00,10.00){\makebox(0,0)[cc]{$0$}}
\put(90.00,10.00){\makebox(0,0)[cc]{$1$}}
\put(100.00,10.00){\makebox(0,0)[cc]{$0$}}
\put(110.00,10.00){\makebox(0,0)[cc]{$1$}}
\put(120.00,10.00){\makebox(0,0)[cc]{$0$}}
\put(130.00,10.00){\makebox(0,0)[cc]{$1$}}
\put(140.00,10.00){\makebox(0,0)[cc]{$0$}}
\put(15.00,00.00){\makebox(0,0)[cc]{$0$}}
\put(25.00,00.00){\makebox(0,0)[cc]{$0$}}
\put(35.00,00.00){\makebox(0,0)[cc]{$1$}}
\put(45.00,00.00){\makebox(0,0)[cc]{$0$}}
\put(55.00,00.00){\makebox(0,0)[cc]{$0$}}
\put(65.00,00.00){\makebox(0,0)[cc]{$0$}}
\put(75.00,00.00){\makebox(0,0)[cc]{$0$}}
\put(85.00,00.00){\makebox(0,0)[cc]{$0$}}
\put(95.00,00.00){\makebox(0,0)[cc]{$0$}}
\put(105.00,00.00){\makebox(0,0)[cc]{$0$}}
\put(115.00,00.00){\makebox(0,0)[cc]{$0$}}
\put(125.00,00.00){\makebox(0,0)[cc]{$0$}}
\put(135.00,00.00){\makebox(0,0)[cc]{$0$}}
\end{picture}
\\[5mm]
On the other hand $|\sigma\rangle$'s have a similar form:

\vspace{5mm}
\unitlength 0.7mm
\linethickness{0.4pt}
\begin{picture}(210.00,10.00)
\put(200.00,5.00){\makebox(0,0)[cc]{$\sigma$.}}
\put(10.00,10.00){\makebox(0,0)[cc]{{\bf H}}}
\put(20.00,10.00){\makebox(0,0)[cc]{$0$}}
\put(30.00,10.00){\makebox(0,0)[cc]{{\bf H}}}
\put(40.00,10.00){\makebox(0,0)[cc]{{\bf H}}}
\put(50.00,10.00){\makebox(0,0)[cc]{{\bf H}}}
\put(60.00,10.00){\makebox(0,0)[cc]{$0$}}
\put(70.00,10.00){\makebox(0,0)[cc]{$0$}}
\put(80.00,10.00){\makebox(0,0)[cc]{{\bf H}}}
\put(90.00,10.00){\makebox(0,0)[cc]{$0$}}
\put(100.00,10.00){\makebox(0,0)[cc]{{\bf H}}}
\put(110.00,10.00){\makebox(0,0)[cc]{{\bf H}}}
\put(120.00,10.00){\makebox(0,0)[cc]{$0$}}
\put(130.00,10.00){\makebox(0,0)[cc]{{\bf H}}}
\put(140.00,10.00){\makebox(0,0)[cc]{{\bf H}}}
\put(15.00,00.00){\makebox(0,0)[cc]{$0$}}
\put(25.00,00.00){\makebox(0,0)[cc]{{\bf H}}}
\put(35.00,00.00){\makebox(0,0)[cc]{$0$}}
\put(45.00,00.00){\makebox(0,0)[cc]{{\bf H}}}
\put(55.00,00.00){\makebox(0,0)[cc]{$0$}}
\put(65.00,00.00){\makebox(0,0)[cc]{$0$}}
\put(75.00,00.00){\makebox(0,0)[cc]{{\bf H}}}
\put(85.00,00.00){\makebox(0,0)[cc]{{\bf H}}}
\put(95.00,00.00){\makebox(0,0)[cc]{{\bf H}}}
\put(105.00,00.00){\makebox(0,0)[cc]{$0$}}
\put(115.00,00.00){\makebox(0,0)[cc]{$0$}}
\put(125.00,00.00){\makebox(0,0)[cc]{$0$}}
\put(135.00,00.00){\makebox(0,0)[cc]{{\bf H}}}
\end{picture}
\\[5mm]
% Recall that the numerical equivalence for
%  ${\bf P}=0$ and ${\bf H}=-1$ monomers.
Where we show {\bf P} monomers by numerical equivalence of them.
Recall that numeric equivalence for {\bf H} monomers is $-1$.
Energy of any sequence in any spatial
configuration is calculated by inner product of its $|\sigma\rangle$
to corresponding $|X\rangle$. For the above $|\sigma\rangle$ and $|X\rangle$
the energy is $5{\bf H}$. This value is related to exact value of
energy according to eq.(\ref{xvec}) by a  factor of two and two
sequence dependent additional terms, since  we are interested in the
   ground state and
the energy gap of a sequence, 
 the sequence dependent term may be ignored,
as structure determines these quantities alone.

By construction any $|X\rangle$ has six $1$'s in odd sites,
and one $1$ in even sites. If we don't consider any other constraint
for $|X\rangle$, we obtain  an upper limit for number of $|X\rangle$'s.
\be
n={14 \choose 6}\times{13 \choose 1}=39039.
\ee
This is far larger than the number of possible $|X\rangle$'s which we
have obtained by enumeration, that is  $6291$.
The fact that all 39039 possible configuration don't exist points to extra
constraints which are yet to be discussed.
If all 39039 of $|X\rangle$'s were to exist each of them would
have to be unique ground state of only one sequence,
 thus removing all interest! To see this, it is enough to
insert  an  {\bf H} into $|X\rangle$ where ever one finds a 1,
and {\bf P} for zeros.
Indeed absence of some of these vectors in real world makes
some of the other more preferable in nature.

The connectivity of a self avoiding walk, further constraints
the  $|X\rangle$. For example to pass through  centre site,
the walk has to pass through two face sites. This means
that the only $1$ (corresponding to centre site) 
in even sites must be sandwiched between two $1$'s
in odd sites (face sites).
This constraint reduces the number of possible $|X\rangle$'s.
Two $1$'s in odd sites are fixed by even $1$, and only 12 sites 
remain for four other $1$'s. Then there are,
\be
n={12 \choose 4}\times{13 \choose 1}=6435,
\ee
vectors. This number is still larger than exact number of 
$|X\rangle$s by  $144$. Although due to our enumeration we know
 these $144$ vectors,
we can not find the complex constraints which  prune them out,
and we shall continue our calculation
as though these 144 vectors were correct.
Of course the values are different from exact enumeration, however 
it can be seen that this difference is not too large, and it may 
be considered as an approximation to the exact solution. 
Also we aid a computer enumeration including the extra 144 vectors and
have compared the results with the combinatorial calculation.
This has served as a check on our code.

We now proceed to calculate  $N'_s$ for  the following  example:

\vspace{5mm}
\unitlength 0.7mm
\linethickness{0.4pt}
\begin{picture}(210.00,20.00)
\put(200.00,10.00){\makebox(0,0)[cc]{$X_0$,}}
\put(10.00,15.00){\makebox(0,0)[cc]{$1$}}
\put(20.00,15.00){\makebox(0,0)[cc]{$0$}}
\put(30.00,15.00){\makebox(0,0)[cc]{$1$}}
\put(40.00,15.00){\makebox(0,0)[cc]{$1$}}
\put(50.00,15.00){\makebox(0,0)[cc]{$1$}}
\put(60.00,15.00){\makebox(0,0)[cc]{$0$}}
\put(70.00,15.00){\makebox(0,0)[cc]{$0$}}
\put(80.00,15.00){\makebox(0,0)[cc]{$0$}}
\put(90.00,15.00){\makebox(0,0)[cc]{$1$}}
\put(100.00,15.00){\makebox(0,0)[cc]{$0$}}
\put(110.00,15.00){\makebox(0,0)[cc]{$0$}}
\put(120.00,15.00){\makebox(0,0)[cc]{$0$}}
\put(130.00,15.00){\makebox(0,0)[cc]{$1$}}
\put(140.00,15.00){\makebox(0,0)[cc]{$0$}}
\put(15.00,5.00){\makebox(0,0)[cc]{$0$}}
\put(25.00,5.00){\makebox(0,0)[cc]{$0$}}
\put(35.00,5.00){\makebox(0,0)[cc]{$1$}}
\put(45.00,5.00){\makebox(0,0)[cc]{$0$}}
\put(55.00,5.00){\makebox(0,0)[cc]{$0$}}
\put(65.00,5.00){\makebox(0,0)[cc]{$0$}}
\put(75.00,5.00){\makebox(0,0)[cc]{$0$}}
\put(85.00,5.00){\makebox(0,0)[cc]{$0$}}
\put(95.00,5.00){\makebox(0,0)[cc]{$0$}}
\put(105.00,5.00){\makebox(0,0)[cc]{$0$}}
\put(115.00,5.00){\makebox(0,0)[cc]{$0$}}
\put(125.00,5.00){\makebox(0,0)[cc]{$0$}}
\put(135.00,5.00){\makebox(0,0)[cc]{$0$}}
\put(25.00,20.00){\line(1,0){20.00}}
\multiput(45.00,20.00)(-0.12,-0.24){84}{\line(0,-1){0.24}}
\multiput(35.00,0.00)(-0.12,0.24){84}{\line(0,1){0.24}}
\end{picture}

First let us introduce some new parameters and notations.
We will show the energy of a $|\sigma\rangle$ in an $|X\rangle$ as:
\be
E=E(a,b,c)=(a+b+c)H,
\ee
where $a,b$ and $c$ are related to energy parts which come from centre 
($1$ in lower row), faces which are connected to centre, 
and energy of other parts, 
respectively. For example energy of following $|\sigma\rangle$:

\vspace{5mm}
\unitlength 0.7mm
\linethickness{0.4pt}
\begin{picture}(210.00,10.00)
\put(200.00,5.00){\makebox(0,0)[cc]{$\sigma$,}}
\put(10.00,10.00){\makebox(0,0)[cc]{{\bf H}}}
\put(20.00,10.00){\makebox(0,0)[cc]{$0$}}
\put(30.00,10.00){\makebox(0,0)[cc]{{\bf H}}}
\put(40.00,10.00){\makebox(0,0)[cc]{{\bf H}}}
\put(50.00,10.00){\makebox(0,0)[cc]{$0$}}
\put(60.00,10.00){\makebox(0,0)[cc]{{\bf H}}}
\put(70.00,10.00){\makebox(0,0)[cc]{$0$}}
\put(80.00,10.00){\makebox(0,0)[cc]{$0$}}
\put(90.00,10.00){\makebox(0,0)[cc]{{\bf H}}}
\put(100.00,10.00){\makebox(0,0)[cc]{{\bf H}}}
\put(110.00,10.00){\makebox(0,0)[cc]{$0$}}
\put(120.00,10.00){\makebox(0,0)[cc]{$0$}}
\put(130.00,10.00){\makebox(0,0)[cc]{{\bf H}}}
\put(140.00,10.00){\makebox(0,0)[cc]{{\bf H}}}
\put(15.00,00.00){\makebox(0,0)[cc]{$0$}}
\put(25.00,00.00){\makebox(0,0)[cc]{{\bf H}}}
\put(35.00,00.00){\makebox(0,0)[cc]{$0$}}
\put(45.00,00.00){\makebox(0,0)[cc]{$0$}}
\put(55.00,00.00){\makebox(0,0)[cc]{{\bf H}}}
\put(65.00,00.00){\makebox(0,0)[cc]{{\bf H}}}
\put(75.00,00.00){\makebox(0,0)[cc]{{\bf H}}}
\put(85.00,00.00){\makebox(0,0)[cc]{$0$}}
\put(95.00,00.00){\makebox(0,0)[cc]{{\bf H}}}
\put(105.00,00.00){\makebox(0,0)[cc]{$0$}}
\put(115.00,00.00){\makebox(0,0)[cc]{$0$}}
\put(125.00,00.00){\makebox(0,0)[cc]{{\bf H}}}
\put(135.00,00.00){\makebox(0,0)[cc]{$0$}}
\end{picture}
\\[5mm]
in $|X_0\rangle$,
is $E(0,2,3)=5{\bf H}$.

Besides, we name the number of pairs of $1$'s in the upper row of
 $|X\rangle$ as $z$ and the number of
$1$'s in two ends of vectors as $y$.
For $|X_0\rangle$, $z=2$, and $y=1$.

Now we try to count the number of all  polymers which have their 
energy  minimised in $|X_0\rangle$
 and, there is no other $|X\rangle$
with energy equal to ground state for them. To do this we discuss 
all possible cases.

{\bf i:} $E(1,2,4)$

Such polymers  have
at least seven {\bf H}  sites corresponding to $1$'s of $X_0$.
These polymers have minimum possible energy, thus $X_0$ is a minimum 
energy configuration for them. But it must be checked whether it is a  
unique ground state  or not. First consider 
polymers which in addition to these seven {\bf H}'s have another {\bf H}
monomer in their upper row sites,

\vspace{5mm}
\unitlength 0.7mm
\linethickness{0.4pt}
\begin{picture}(210.00,20.00)
\put(200.00,10.00){\makebox(0,0)[cc]{$\sigma_1$.}}
\put(10.00,15.00){\makebox(0,0)[cc]{{\bf H}}}
\put(20.00,15.00){\makebox(0,0)[cc]{$0$}}
\put(30.00,15.00){\makebox(0,0)[cc]{{\bf H}}}
\put(40.00,15.00){\makebox(0,0)[cc]{{\bf H}}}
\put(50.00,15.00){\makebox(0,0)[cc]{{\bf H}}}
\put(60.00,15.00){\makebox(0,0)[cc]{$0$}}
\put(70.00,15.00){\makebox(0,0)[cc]{$0$}}
\put(80.00,15.00){\makebox(0,0)[cc]{$0$}}
\put(90.00,15.00){\makebox(0,0)[cc]{{\bf H}}}
\put(100.00,15.00){\makebox(0,0)[cc]{$0$}}
\put(110.00,15.00){\makebox(0,0)[cc]{{\bf H}}}
\put(120.00,15.00){\makebox(0,0)[cc]{$0$}}
\put(130.00,15.00){\makebox(0,0)[cc]{{\bf H}}}
\put(140.00,15.00){\makebox(0,0)[cc]{$0$}}
\put(15.00,5.00){\makebox(0,0)[cc]{$0$}}
\put(25.00,5.00){\makebox(0,0)[cc]{$0$}}
\put(35.00,5.00){\makebox(0,0)[cc]{{\bf H}}}
\put(45.00,5.00){\makebox(0,0)[cc]{$0$}}
\put(55.00,5.00){\makebox(0,0)[cc]{$0$}}
\put(65.00,5.00){\makebox(0,0)[cc]{$0$}}
\put(75.00,5.00){\makebox(0,0)[cc]{$0$}}
\put(85.00,5.00){\makebox(0,0)[cc]{$0$}}
\put(95.00,5.00){\makebox(0,0)[cc]{$0$}}
\put(105.00,5.00){\makebox(0,0)[cc]{$0$}}
\put(115.00,5.00){\makebox(0,0)[cc]{$0$}}
\put(125.00,5.00){\makebox(0,0)[cc]{$0$}}
\put(135.00,5.00){\makebox(0,0)[cc]{$0$}}
\end{picture}
\\
The energy of this sequence in following $|X\rangle$ is $7{\bf H}$ too.

\vspace{5mm}
\unitlength 0.7mm
\linethickness{0.4pt}
\begin{picture}(210.00,20.00)
\put(200.00,10.00){\makebox(0,0)[cc]{$X_1$.}}
\put(10.00,15.00){\makebox(0,0)[cc]{$1$}}
\put(20.00,15.00){\makebox(0,0)[cc]{$0$}}
\put(30.00,15.00){\makebox(0,0)[cc]{$1$}}
\put(40.00,15.00){\makebox(0,0)[cc]{$1$}}
\put(50.00,15.00){\makebox(0,0)[cc]{$1$}}
\put(60.00,15.00){\makebox(0,0)[cc]{$0$}}
\put(70.00,15.00){\makebox(0,0)[cc]{$0$}}
\put(80.00,15.00){\makebox(0,0)[cc]{$0$}}
\put(90.00,15.00){\makebox(0,0)[cc]{$1$}}
\put(100.00,15.00){\makebox(0,0)[cc]{$0$}}
\put(110.00,15.00){\makebox(0,0)[cc]{$1$}}
\put(120.00,15.00){\makebox(0,0)[cc]{$0$}}
\put(130.00,15.00){\makebox(0,0)[cc]{$0$}}
\put(140.00,15.00){\makebox(0,0)[cc]{$0$}}
\put(15.00,5.00){\makebox(0,0)[cc]{$0$}}
\put(25.00,5.00){\makebox(0,0)[cc]{$0$}}
\put(35.00,5.00){\makebox(0,0)[cc]{$1$}}
\put(45.00,5.00){\makebox(0,0)[cc]{$0$}}
\put(55.00,5.00){\makebox(0,0)[cc]{$0$}}
\put(65.00,5.00){\makebox(0,0)[cc]{$0$}}
\put(75.00,5.00){\makebox(0,0)[cc]{$0$}}
\put(85.00,5.00){\makebox(0,0)[cc]{$0$}}
\put(95.00,5.00){\makebox(0,0)[cc]{$0$}}
\put(105.00,5.00){\makebox(0,0)[cc]{$0$}}
\put(115.00,5.00){\makebox(0,0)[cc]{$0$}}
\put(125.00,5.00){\makebox(0,0)[cc]{$0$}}
\put(135.00,5.00){\makebox(0,0)[cc]{$0$}}
\put(25.00,20.00){\line(1,0){20.00}}
\multiput(45.00,20.00)(-0.12,-0.24){84}{\line(0,-1){0.24}}
\multiput(35.00,0.00)(-0.12,0.24){84}{\line(0,1){0.24}}
\end{picture}
\\[5mm]
Then the ground state of  polymers which have 
additional {\bf H} monomers in corresponding to 
upper row $0$'s of $X_0$, is degenerate, and they don't count in $N'_s$ 
of $X_0$. The above discussion is independent of value of
$a$ and $b$ in $E(a,b,4)$, and degrees of freedom to choose
sites for {\bf H} monomers is limited to lower row sites.

For the $|X\rangle$ with $z\neq 1$ (like $X_0$) polymers can not
have {\bf H} monomers in the lower sites  between  two upper  
row $1$'s.
For example, the following sequence,

\vspace{5mm}
\unitlength 0.7mm
\linethickness{0.4pt}
\begin{picture}(210.00,20.00)
\put(200.00,10.00){\makebox(0,0)[cc]{$\sigma_2$,}}
\put(10.00,15.00){\makebox(0,0)[cc]{{\bf H}}}
\put(20.00,15.00){\makebox(0,0)[cc]{$0$}}
\put(30.00,15.00){\makebox(0,0)[cc]{{\bf H}}}
\put(40.00,15.00){\makebox(0,0)[cc]{{\bf H}}}
\put(50.00,15.00){\makebox(0,0)[cc]{{\bf H}}}
\put(60.00,15.00){\makebox(0,0)[cc]{$0$}}
\put(70.00,15.00){\makebox(0,0)[cc]{$0$}}
\put(80.00,15.00){\makebox(0,0)[cc]{$0$}}
\put(90.00,15.00){\makebox(0,0)[cc]{{\bf H}}}
\put(100.00,15.00){\makebox(0,0)[cc]{$0$}}
\put(110.00,15.00){\makebox(0,0)[cc]{$0$}}
\put(120.00,15.00){\makebox(0,0)[cc]{$0$}}
\put(130.00,15.00){\makebox(0,0)[cc]{{\bf H}}}
\put(140.00,15.00){\makebox(0,0)[cc]{$0$}}
\put(15.00,5.00){\makebox(0,0)[cc]{$0$}}
\put(25.00,5.00){\makebox(0,0)[cc]{$0$}}
\put(35.00,5.00){\makebox(0,0)[cc]{{\bf H}}}
\put(45.00,5.00){\makebox(0,0)[cc]{{\bf H}}}
\put(55.00,5.00){\makebox(0,0)[cc]{$0$}}
\put(65.00,5.00){\makebox(0,0)[cc]{$0$}}
\put(75.00,5.00){\makebox(0,0)[cc]{$0$}}
\put(85.00,5.00){\makebox(0,0)[cc]{$0$}}
\put(95.00,5.00){\makebox(0,0)[cc]{$0$}}
\put(105.00,5.00){\makebox(0,0)[cc]{$0$}}
\put(115.00,5.00){\makebox(0,0)[cc]{$0$}}
\put(125.00,5.00){\makebox(0,0)[cc]{$0$}}
\put(135.00,5.00){\makebox(0,0)[cc]{$0$}}
\end{picture}
\\[5mm]
has energy $7{\bf H}$ in following $|X\rangle$ too.

\vspace{5mm}
\unitlength 0.7mm
\linethickness{0.4pt}
\begin{picture}(210.00,20.00)
\put(200.00,10.00){\makebox(0,0)[cc]{$X_2$.}}
\put(10.00,15.00){\makebox(0,0)[cc]{$1$}}
\put(20.00,15.00){\makebox(0,0)[cc]{$0$}}
\put(30.00,15.00){\makebox(0,0)[cc]{$1$}}
\put(40.00,15.00){\makebox(0,0)[cc]{$1$}}
\put(50.00,15.00){\makebox(0,0)[cc]{$1$}}
\put(60.00,15.00){\makebox(0,0)[cc]{$0$}}
\put(70.00,15.00){\makebox(0,0)[cc]{$0$}}
\put(80.00,15.00){\makebox(0,0)[cc]{$0$}}
\put(90.00,15.00){\makebox(0,0)[cc]{$1$}}
\put(100.00,15.00){\makebox(0,0)[cc]{$0$}}
\put(110.00,15.00){\makebox(0,0)[cc]{$0$}}
\put(120.00,15.00){\makebox(0,0)[cc]{$0$}}
\put(130.00,15.00){\makebox(0,0)[cc]{$1$}}
\put(140.00,15.00){\makebox(0,0)[cc]{$0$}}
\put(15.00,5.00){\makebox(0,0)[cc]{$0$}}
\put(25.00,5.00){\makebox(0,0)[cc]{$0$}}
\put(35.00,5.00){\makebox(0,0)[cc]{$0$}}
\put(45.00,5.00){\makebox(0,0)[cc]{$1$}}
\put(55.00,5.00){\makebox(0,0)[cc]{$0$}}
\put(65.00,5.00){\makebox(0,0)[cc]{$0$}}
\put(75.00,5.00){\makebox(0,0)[cc]{$0$}}
\put(85.00,5.00){\makebox(0,0)[cc]{$0$}}
\put(95.00,5.00){\makebox(0,0)[cc]{$0$}}
\put(105.00,5.00){\makebox(0,0)[cc]{$0$}}
\put(115.00,5.00){\makebox(0,0)[cc]{$0$}}
\put(125.00,5.00){\makebox(0,0)[cc]{$0$}}
\put(135.00,5.00){\makebox(0,0)[cc]{$0$}}
\put(35.00,20.00){\line(1,0){20.00}}
\multiput(55.00,20.00)(-0.12,-0.24){84}{\line(0,-1){0.24}}
\multiput(45.00,0.00)(-0.12,0.24){84}{\line(0,1){0.24}}
\end{picture}
\\[5mm]
Then the contribution of polymers with $E(1,2,4)$ in $N'_s$ is:

\be
 N'_s({\bf i})=2^{12-(z-1)}=2^{13-z}
\ee

{\bf ii:} $E(0,2,4)$  

In this case if $z > 1$ (such as $X_0$) the ground state is degenerate.
It can be seen that any sequence  with energy $E=(0,2,4)$ in $X_0$
state has the same energy in $X_2$ state. In the case $z=1$,
only the sites in lower row by condition that they are not a neighbour 
of corresponding upper $1$'s of $X$, have freedom to be an {\bf H} or
{\bf P} monomer. There are $2 \times 6 -z-y$ sites which don't have 
this freedom in lower row. Then,
\be
N'_s({\bf ii})= \left\{ \begin{array}{ll}
		     2^{2-y}   & \ \ \ \ \ z=1 \\
		     0         & \ \ \ \ \  z>1 \\
		\end{array}
     \right.
\ee

{\bf iii:} $E(1,0,4)$  

In this case there is only one sequence with nondegenerate 
ground state. For our example, $X_0$, this sequence is,

\vspace{5mm}
\unitlength 0.7mm
\linethickness{0.4pt}
\begin{picture}(210.00,20.00)
\put(200.00,10.00){\makebox(0,0)[cc]{$\sigma_3$.}}
\put(10.00,15.00){\makebox(0,0)[cc]{{\bf H}}}
\put(20.00,15.00){\makebox(0,0)[cc]{$0$}}
\put(30.00,15.00){\makebox(0,0)[cc]{$0$}}
\put(40.00,15.00){\makebox(0,0)[cc]{$0$}}
\put(50.00,15.00){\makebox(0,0)[cc]{{\bf H}}}
\put(60.00,15.00){\makebox(0,0)[cc]{$0$}}
\put(70.00,15.00){\makebox(0,0)[cc]{$0$}}
\put(80.00,15.00){\makebox(0,0)[cc]{$0$}}
\put(90.00,15.00){\makebox(0,0)[cc]{{\bf H}}}
\put(100.00,15.00){\makebox(0,0)[cc]{$0$}}
\put(110.00,15.00){\makebox(0,0)[cc]{$0$}}
\put(120.00,15.00){\makebox(0,0)[cc]{$0$}}
\put(130.00,15.00){\makebox(0,0)[cc]{{\bf H}}}
\put(140.00,15.00){\makebox(0,0)[cc]{$0$}}
\put(15.00,5.00){\makebox(0,0)[cc]{$0$}}
\put(25.00,5.00){\makebox(0,0)[cc]{$0$}}
\put(35.00,5.00){\makebox(0,0)[cc]{{\bf H}}}
\put(45.00,5.00){\makebox(0,0)[cc]{$0$}}
\put(55.00,5.00){\makebox(0,0)[cc]{$0$}}
\put(65.00,5.00){\makebox(0,0)[cc]{$0$}}
\put(75.00,5.00){\makebox(0,0)[cc]{$0$}}
\put(85.00,5.00){\makebox(0,0)[cc]{$0$}}
\put(95.00,5.00){\makebox(0,0)[cc]{$0$}}
\put(105.00,5.00){\makebox(0,0)[cc]{$0$}}
\put(115.00,5.00){\makebox(0,0)[cc]{$0$}}
\put(125.00,5.00){\makebox(0,0)[cc]{$0$}}
\put(135.00,5.00){\makebox(0,0)[cc]{$0$}}
\end{picture}

In the above sequence 
changing any {\bf P} monomer to {\bf H} type, 
will cause the ground state to becomes degenerate. Then,
\be 
N'_s({\bf iii})=1
\ee

{\bf iv:} $E(1,1,4)$  

For this case $b$ is $1$, and if this $1$ comes from right or left 
neighbour of lower $1$, it has different solutions.  
Then we introduce new parameters ($z_R$, $y_R$) and ($z_L$, $y_L$),
 which are
similar to old $z$ and $y$, when right or left neighbour $1$ of lower $1$
will be omitted. For $X_0$ we have $z_R=0$, $z_L=1$ and $y_R=y_L=y=1$.
By introducing this new parameters 
this case is very similar to case {\bf ii}, and the difference 
comes from number of corresponding $1$'s in upper row (five instead six),
and no restriction in value of $z$.
Then,

\be
N'_s({\bf iv})=2^{3+z_R+y_R}+2^{3+z_L+y_L}.
\ee

{\bf v:} Other cases 

All of the other cases for ground state energy are degenerate,
and need not be considered.

With this analysis it is possible to find $N'_s$ for any $|X\rangle$.
For our $X_0$ example it is,
\bea
N'_s({\bf i})   &=& 2^{11}  \nonumber  \\
N'_s({\bf ii})  &=& 0  \nonumber  \\
N'_s({\bf iii}) &=& 1  \nonumber  \\
N'_s({\bf iv}) &=& 2^{4} + 2^5 ,\nonumber  
\eea
that gives,
\bea
N'_s(X_0)= 2097   \nonumber 
\eea

In this way all of the values of $N'_s$'s can be calculated.
Had the $144$ additional structures been taken out, the calculation
of $N'_s$ for the problem would correspond to enumeration exactly.
However taking these structures out is too complex and would have
to be done case by case.
Besides of the value of $N'_s$, this calculation
shows that the sequences whit non-degenerate ground state
have between 4 to 6 {\bf H} type  monomers in face sites and no one
in corner sites.
Indeed in our model the stability of polymers
is very sensitive to mutation in corner sites.

%%%%%%%%%%%%%%%%%%%%%%%%%%%%%%%%%%%%%%%%%%%%%%%%%%%%%%%%%%%%%%%
\newsection{Nonadditive potentials}

In the case $\gamma \neq 0$ the potential is non-additive.
In this case we can write the energy of $\alpha$th sequence
in $\mu$th spatial
configuration as:
\be
E_{\alpha\mu}=\langle \sigma_\alpha|P_\mu\rangle
-{\textstyle \frac{1}{2}}\gamma \langle \sigma_\alpha|M_\mu|\sigma_\alpha\rangle.
\label{xm}
\ee
Where $\sigma$ and $P$ are the sequence and neighbourhood vectors,
that introduced in previous sections. $M$ is the
adjacency matrix for this configuration.
\be
M_{ij}= \left\{ \begin{array}{llr}
		     1   & \ \ \ \ \ {\rm if \ the}\ i{\rm th \ and}\ j{\rm th\
		     monomers\ are\ adjacent;} \\
		     0   & \ \ \ \ \  {\rm otherwise}. \\
		\end{array}
     \right.
\ee

Any $|P\rangle$ has $N_d$ different $M$-matrices. 
The first term in eq. (\ref{xm}) was calculated in the case $\gamma =0$,
and we need calculate only the last part.
The aim of our calculation is to find the  ground state.
In any compact configuration in a $3\times3\times3$ cube, there are $28$
non-sequential neighbour pairs. Thus the contribution of the last term
in energy is less than $28\gamma$. We have shown that energy spectrum
for the previous case has a ladder structure with energy gap equal  to $2$.
In this case these split to some sublevels (fig. 7).
Then if we choose $\gamma < \frac{2}{28}$ the levels are separate.
Of course this is a lower estimation for $\gamma$. In the next
section we will obtain a better estimate for
lower and upper limits of the critical value of $\gamma$.
\bfig
\bcen
\input{fig7.tex}
\ecen
\caption{
	       Energy levels of additive potential split to sublevels 
	       for non-additive potential.}

\efig

From the result of the additive potential we have a subset $P$ in the
 space of all spatial configurations which gives the minimum energy to
folding. This $P$ subset has $N_d$ members which all of them have the
same $|P\rangle$.
For small $\gamma$'s the ground state and the first excited state are between
these $N_d$ structures, and it is not necessary
to calculate the energy for all of $103,346$ spatial structures for any
sequence,except for sequences which their
ground state is in structures with $N_d=1$.
For The $N_d=1$ structures the value of $N_s$ does not change, and
it is not necessary to run the program.
The first
 excited state of these sequences are in an other $P$ subset.
Thus to find the
energy gap for them the program must
be run over all of the $103,346$ structures.
We have calculated this energy spectrum, and
 have found the new $N_s$ for all $103,346$ structures. We show the results
for $51,704$ configuration which are unrelated by reverse labelling
symmetry  in fig. 8. We have found the energy gap for first excited state
for all sequences.
 You can see the diagram of mean of energy gap {\it vs.} $N_s$ in
fig. 9. This figure shows that highly designable structures
which are related to
$P$ subsets with one member.
\bfig
\bcen
\epsfxsize=10.0cm
\epsfbox{fig8.eps}
\ecen
\caption{
	       Histogram of $N_s$ for non-additive potential.}

\efig
\bfig
\bcen
\epsfxsize=10.0cm
\epsfbox{fig9.eps}
\ecen
\caption{
	       The mean of energy gap {\it vs.} $N_s$. 
	       There is a jump in energy gap for highly designable
	       structures. All of these highly designable structures 
	       have $N_d=1$.}

\efig

In this enumeration we have calculated the energy spectrum for all
of the sequences which have nondegenerate ground state for
the additive potential.
We had removed some of the  sequences because of degeneracy of
ground state in  additive potential case.
It is possible that this degeneracy will be  removed
by the non-additive part of the potential, 
 and some of the sequences have unique ground
 state for non-additive potential.
But the energy gap for these sequences is of order of $\gamma$,
and if we consider them  it causes  a shift in horizontal axes to
bigger $N_s$ and bring down the points nearer to $\gamma$ value
in vertical direction in fig. 9.
These make this figure more similar to results of Li {\it et al.}
\cite{Li}.
In their work  the energy gap for
 low designable structures are of order of $\gamma$ ( they choose
 $\gamma = 0.3$) also.

%%%%%%%%%%%%%%%%%%%%%%%%%%%%%%%%%%%%%%%%%%%%%%%%%%%%%%%%%%%%%%
\newsection{Estimation of  $\gamma_c$}

The energy levels for the additive potential have a ladder structure,
as it had been proven in previous sections.
The energy gaps between the levels is $2$ in our arbitrary energy unit.

In the case  of $\gamma \neq 0$ the energy has two parts (eq. \ref{xm}).
The first part comes from additive part of potential and does not change.
Second part comes from non-additive part of potential, and is equal to
number of {\bf H-H} non-sequential neighbours in spatial configuration.
Because this non-additive part, the energy spectrum is changed,
and any level is splited to some sublevels (fig. 7).

If contribution of the second part to energy is less than $2$,
for all structures,  the ground state and
excited state of any polymer is between $|P\rangle$ partners
of its  ground state for additive potential ,
except for $|P\rangle$ with $N_d=1$, where there is only ground state.

Let $\delta e_0$ be the difference of ground state energies
of additive and non-additive potentials,
and $\delta e_1$ be difference  energy of first excited state in the
case of $\gamma=0$ with minimum of new energies for the sequence in the
structures corresponding to these excited states
(there is no uniqueness constraint for excited states).
If $\delta e_0 - \delta e_1 < 2$ the ground state doesn't
change and the values of $N_s$ for structures that we presented in
past section don't change.
By increasing $\gamma$,  the absolute values of
$\delta e_0$ and $\delta e_1$ increase.

To find the difference between $\delta e_0$ and $\delta e_1$ one have to
calculate the difference in {\bf H-H} contacts in ground state structure
and maximum of {\bf H-H} contacts in excited level structures.

This difference has two sources.
Because the energy levels in the case $\gamma=0$ are separated
by 2, then difference of them comes from replacing a {\bf H}
monomer from $O$ site to an $L$ site, or from an $F$ site to a $C$ site.
Both of them cause increasing in energy by 2.
But it is possible that these replacing  decrease the energy by $2\gamma$.
For example consider one $F$ site with no {\bf H}
neighbour will go to one $C$ site with two non-sequential {\bf H}
neighbours (this monomer must be an end residue).
then this gives an upper limit for $\gamma_c$, that is $1$.

The other source for increasing the {\bf H-H} contacts,
comes from replacing {\bf H} monomers in  $L$ and $F$ sites
by the same type sites. These changes only  are relevant in
the case $\gamma \neq 0$. The maximum of increasing in {\bf H-H}
contacts because these replacing are $6\gamma$, related to the
sequences which have 4 {\bf H} monomers in the $F$ sites and 5 to 7 in
$L$ sites in their ground state structures. Thus the lower limit for
$\gamma_c$ is $\frac{2}{2+6}=0.25$. Therefore,
\be
0.25 < \gamma_c < 1.
\ee
This shows that there is a non zero  value for $\gamma_c$,
which for $\gamma$ less  than it, the ground state
structure of sequences doesn't change.
Indeed  $\gamma_c$ distinguishes two phases.
If $\gamma < \gamma_c$, the degree of designability of structures
is independent of $\gamma$, and the change in value of $\gamma$
only changes the energy gaps.
On the other hand for $\gamma > \gamma_c$,
the designability of structures becomes  sensitive to the value of
$\gamma$, and the patterns of highly designable structures will be
changed if the potential changes.

If  the designability is the answer of ``why has the nature
selected a small fraction of possible configurations for
folded states?", the above discussion  shows that this
selection is potential independent if $\gamma < \gamma_c$,
and sensitive to inter monomers interactions if $\gamma >\gamma_c$.

{\bf Acknowledgements}

We would like to thank J. Davoudi for motivating the problem,
 R. Golestanian and S. Saber  for helpful comments,
 and S. Rouhani for helpful comments throughout the work and reading
 the manuscript.

%%%%%%%%%%%%%%%%%%%%%%%%%%%%%%%%%%%%%%%%%%%%%%%%%%%%%%%%%%%%%%%
%\newsection{Discussion}
%%%%%%%%%%%%%%%%%%%%%%%%%%%%%%%%%%%%%%%%%%%%%%%%%%%%%%%%%%%%%%%

%%%%%%%  References  %%%%%%%%%%%%%%%%%%%%%%%%%%%%%%%%%%%%%%%
%%\vspace{5mm}
%\newpage
%
\newcommand{\PNAS}[1]{ Pros.\ Natl.\ Acad.\ Sci.\ USA\ {\bf #1}}
\newcommand{\JCP}[1]{ J.\ Chem.\ Phys.\ {\bf #1}}

%%%%%%%%%%%%%%%%%%%%%%%%%%%
\end{document}

%% file: fig1.tex
%TexCad Options
%\grade{\off}
%\emlines{\off}
%\beziermacro{\on}
%\reduce{\on}
%\snapping{\off}
%\quality{2.00}
%\graddiff{0.01}
%\snapasp{1}
%\zoom{2.98}
\unitlength 1.00mm
\linethickness{0.8pt}
\begin{picture}(117.28,67.97)
\put(16.33,65.00){\line(1,0){2.00}}
\put(20.33,65.00){\line(1,0){2.00}}
\put(24.33,65.00){\line(1,0){2.00}}
\put(28.33,65.00){\line(1,0){2.00}}
\put(32.33,65.00){\line(1,0){2.00}}
\put(36.33,65.00){\line(1,0){2.00}}
\put(40.33,65.00){\line(1,0){2.00}}
\put(44.33,65.00){\line(1,0){2.00}}
\put(48.33,65.00){\line(1,0){2.00}}
\put(52.33,65.00){\line(1,0){2.00}}
\put(16.33,55.00){\line(1,0){2.00}}
\put(16.33,45.00){\line(1,0){2.00}}
\put(16.33,35.00){\line(1,0){2.00}}
\put(16.33,25.00){\line(1,0){2.00}}
\put(20.33,55.00){\line(1,0){2.00}}
\put(20.33,45.00){\line(1,0){2.00}}
\put(20.33,35.00){\line(1,0){2.00}}
\put(20.33,25.00){\line(1,0){2.00}}
\put(24.33,55.00){\line(1,0){2.00}}
\put(24.33,45.00){\line(1,0){2.00}}
\put(24.33,35.00){\line(1,0){2.00}}
\put(24.33,25.00){\line(1,0){2.00}}
\put(28.33,55.00){\line(1,0){2.00}}
\put(28.33,45.00){\line(1,0){2.00}}
\put(28.33,35.00){\line(1,0){2.00}}
\put(28.33,25.00){\line(1,0){2.00}}
\put(32.33,55.00){\line(1,0){2.00}}
\put(32.33,45.00){\line(1,0){2.00}}
\put(32.33,35.00){\line(1,0){2.00}}
\put(32.33,25.00){\line(1,0){2.00}}
\put(36.33,55.00){\line(1,0){2.00}}
\put(36.33,45.00){\line(1,0){2.00}}
\put(36.33,35.00){\line(1,0){2.00}}
\put(36.33,25.00){\line(1,0){2.00}}
\put(40.33,55.00){\line(1,0){2.00}}
\put(40.33,45.00){\line(1,0){2.00}}
\put(40.33,35.00){\line(1,0){2.00}}
\put(40.33,25.00){\line(1,0){2.00}}
\put(44.33,55.00){\line(1,0){2.00}}
\put(44.33,45.00){\line(1,0){2.00}}
\put(44.33,35.00){\line(1,0){2.00}}
\put(44.33,25.00){\line(1,0){2.00}}
\put(48.33,55.00){\line(1,0){2.00}}
\put(48.33,45.00){\line(1,0){2.00}}
\put(48.33,35.00){\line(1,0){2.00}}
\put(48.33,25.00){\line(1,0){2.00}}
\put(52.33,55.00){\line(1,0){2.00}}
\put(52.33,45.00){\line(1,0){2.00}}
\put(52.33,35.00){\line(1,0){2.00}}
\put(52.33,25.00){\line(1,0){2.00}}
\put(56.33,65.00){\line(0,-1){2.00}}
\put(56.33,61.00){\line(0,-1){2.00}}
\put(56.33,57.00){\line(0,-1){2.00}}
\put(56.33,53.00){\line(0,-1){2.00}}
\put(56.33,49.00){\line(0,-1){2.00}}
\put(56.33,45.00){\line(0,-1){2.00}}
\put(56.33,41.00){\line(0,-1){2.00}}
\put(56.33,37.00){\line(0,-1){2.00}}
\put(56.33,33.00){\line(0,-1){2.00}}
\put(56.33,29.00){\line(0,-1){2.00}}
\put(46.33,65.00){\line(0,-1){2.00}}
\put(36.33,65.00){\line(0,-1){2.00}}
\put(26.33,65.00){\line(0,-1){2.00}}
\put(16.33,65.00){\line(0,-1){2.00}}
\put(46.33,61.00){\line(0,-1){2.00}}
\put(36.33,61.00){\line(0,-1){2.00}}
\put(26.33,61.00){\line(0,-1){2.00}}
\put(16.33,61.00){\line(0,-1){2.00}}
\put(46.33,57.00){\line(0,-1){2.00}}
\put(36.33,57.00){\line(0,-1){2.00}}
\put(26.33,57.00){\line(0,-1){2.00}}
\put(16.33,57.00){\line(0,-1){2.00}}
\put(46.33,53.00){\line(0,-1){2.00}}
\put(36.33,53.00){\line(0,-1){2.00}}
\put(26.33,53.00){\line(0,-1){2.00}}
\put(16.33,53.00){\line(0,-1){2.00}}
\put(46.33,49.00){\line(0,-1){2.00}}
\put(36.33,49.00){\line(0,-1){2.00}}
\put(26.33,49.00){\line(0,-1){2.00}}
\put(16.33,49.00){\line(0,-1){2.00}}
\put(46.33,45.00){\line(0,-1){2.00}}
\put(36.33,45.00){\line(0,-1){2.00}}
\put(26.33,45.00){\line(0,-1){2.00}}
\put(16.33,45.00){\line(0,-1){2.00}}
\put(46.33,41.00){\line(0,-1){2.00}}
\put(36.33,41.00){\line(0,-1){2.00}}
\put(26.33,41.00){\line(0,-1){2.00}}
\put(16.33,41.00){\line(0,-1){2.00}}
\put(46.33,37.00){\line(0,-1){2.00}}
\put(36.33,37.00){\line(0,-1){2.00}}
\put(26.33,37.00){\line(0,-1){2.00}}
\put(16.33,37.00){\line(0,-1){2.00}}
\put(46.33,33.00){\line(0,-1){2.00}}
\put(36.33,33.00){\line(0,-1){2.00}}
\put(26.33,33.00){\line(0,-1){2.00}}
\put(16.33,33.00){\line(0,-1){2.00}}
\put(46.33,29.00){\line(0,-1){2.00}}
\put(36.33,29.00){\line(0,-1){2.00}}
\put(26.33,29.00){\line(0,-1){2.00}}
\put(16.33,29.00){\line(0,-1){2.00}}
\put(76.33,65.00){\line(1,0){2.00}}
\put(80.33,65.00){\line(1,0){2.00}}
\put(84.33,65.00){\line(1,0){2.00}}
\put(88.33,65.00){\line(1,0){2.00}}
\put(92.33,65.00){\line(1,0){2.00}}
\put(96.33,65.00){\line(1,0){2.00}}
\put(100.33,65.00){\line(1,0){2.00}}
\put(104.33,65.00){\line(1,0){2.00}}
\put(108.33,65.00){\line(1,0){2.00}}
\put(112.33,65.00){\line(1,0){2.00}}
\put(76.33,55.00){\line(1,0){2.00}}
\put(76.33,45.00){\line(1,0){2.00}}
\put(76.33,35.00){\line(1,0){2.00}}
\put(76.33,25.00){\line(1,0){2.00}}
\put(80.33,55.00){\line(1,0){2.00}}
\put(80.33,45.00){\line(1,0){2.00}}
\put(80.33,35.00){\line(1,0){2.00}}
\put(80.33,25.00){\line(1,0){2.00}}
\put(84.33,55.00){\line(1,0){2.00}}
\put(84.33,45.00){\line(1,0){2.00}}
\put(84.33,35.00){\line(1,0){2.00}}
\put(84.33,25.00){\line(1,0){2.00}}
\put(88.33,55.00){\line(1,0){2.00}}
\put(88.33,45.00){\line(1,0){2.00}}
\put(88.33,35.00){\line(1,0){2.00}}
\put(88.33,25.00){\line(1,0){2.00}}
\put(92.33,55.00){\line(1,0){2.00}}
\put(92.33,45.00){\line(1,0){2.00}}
\put(92.33,35.00){\line(1,0){2.00}}
\put(92.33,25.00){\line(1,0){2.00}}
\put(96.33,55.00){\line(1,0){2.00}}
\put(96.33,45.00){\line(1,0){2.00}}
\put(96.33,35.00){\line(1,0){2.00}}
\put(96.33,25.00){\line(1,0){2.00}}
\put(100.33,55.00){\line(1,0){2.00}}
\put(100.33,45.00){\line(1,0){2.00}}
\put(100.33,35.00){\line(1,0){2.00}}
\put(100.33,25.00){\line(1,0){2.00}}
\put(104.33,55.00){\line(1,0){2.00}}
\put(104.33,45.00){\line(1,0){2.00}}
\put(104.33,35.00){\line(1,0){2.00}}
\put(104.33,25.00){\line(1,0){2.00}}
\put(108.33,55.00){\line(1,0){2.00}}
\put(108.33,45.00){\line(1,0){2.00}}
\put(108.33,35.00){\line(1,0){2.00}}
\put(108.33,25.00){\line(1,0){2.00}}
\put(112.33,55.00){\line(1,0){2.00}}
\put(112.33,45.00){\line(1,0){2.00}}
\put(112.33,35.00){\line(1,0){2.00}}
\put(112.33,25.00){\line(1,0){2.00}}
\put(116.33,65.00){\line(0,-1){2.00}}
\put(116.33,61.00){\line(0,-1){2.00}}
\put(116.33,57.00){\line(0,-1){2.00}}
\put(116.33,53.00){\line(0,-1){2.00}}
\put(116.33,49.00){\line(0,-1){2.00}}
\put(116.33,45.00){\line(0,-1){2.00}}
\put(116.33,41.00){\line(0,-1){2.00}}
\put(116.33,37.00){\line(0,-1){2.00}}
\put(116.33,33.00){\line(0,-1){2.00}}
\put(116.33,29.00){\line(0,-1){2.00}}
\put(106.33,65.00){\line(0,-1){2.00}}
\put(96.33,65.00){\line(0,-1){2.00}}
\put(86.33,65.00){\line(0,-1){2.00}}
\put(76.33,65.00){\line(0,-1){2.00}}
\put(106.33,61.00){\line(0,-1){2.00}}
\put(96.33,61.00){\line(0,-1){2.00}}
\put(86.33,61.00){\line(0,-1){2.00}}
\put(76.33,61.00){\line(0,-1){2.00}}
\put(106.33,57.00){\line(0,-1){2.00}}
\put(96.33,57.00){\line(0,-1){2.00}}
\put(86.33,57.00){\line(0,-1){2.00}}
\put(76.33,57.00){\line(0,-1){2.00}}
\put(106.33,53.00){\line(0,-1){2.00}}
\put(96.33,53.00){\line(0,-1){2.00}}
\put(86.33,53.00){\line(0,-1){2.00}}
\put(76.33,53.00){\line(0,-1){2.00}}
\put(106.33,49.00){\line(0,-1){2.00}}
\put(96.33,49.00){\line(0,-1){2.00}}
\put(86.33,49.00){\line(0,-1){2.00}}
\put(76.33,49.00){\line(0,-1){2.00}}
\put(106.33,45.00){\line(0,-1){2.00}}
\put(96.33,45.00){\line(0,-1){2.00}}
\put(86.33,45.00){\line(0,-1){2.00}}
\put(76.33,45.00){\line(0,-1){2.00}}
\put(106.33,41.00){\line(0,-1){2.00}}
\put(96.33,41.00){\line(0,-1){2.00}}
\put(86.33,41.00){\line(0,-1){2.00}}
\put(76.33,41.00){\line(0,-1){2.00}}
\put(106.33,37.00){\line(0,-1){2.00}}
\put(96.33,37.00){\line(0,-1){2.00}}
\put(86.33,37.00){\line(0,-1){2.00}}
\put(76.33,37.00){\line(0,-1){2.00}}
\put(106.33,33.00){\line(0,-1){2.00}}
\put(96.33,33.00){\line(0,-1){2.00}}
\put(86.33,33.00){\line(0,-1){2.00}}
\put(76.33,33.00){\line(0,-1){2.00}}
\put(106.33,29.00){\line(0,-1){2.00}}
\put(96.33,29.00){\line(0,-1){2.00}}
\put(86.33,29.00){\line(0,-1){2.00}}
\put(76.33,29.00){\line(0,-1){2.00}}
\put(26.66,55.00){\line(1,0){10.00}}
\put(36.66,65.00){\line(-1,0){20.33}}
\put(16.33,65.00){\line(0,-1){40.00}}
\put(16.33,25.00){\line(1,0){20.00}}
\put(36.33,25.00){\line(0,1){20.00}}
\put(36.33,45.00){\line(1,0){10.00}}
\put(46.33,45.00){\line(0,1){19.67}}
\put(56.33,64.67){\line(0,-1){39.67}}
\put(56.33,25.00){\line(-1,0){9.67}}
\put(86.33,35.00){\line(0,1){20.00}}
\put(86.33,55.00){\line(1,0){10.00}}
\put(96.33,55.00){\line(0,1){10.00}}
\put(96.33,65.00){\line(-1,0){20.00}}
\put(76.33,65.00){\line(0,-1){40.00}}
\put(76.33,25.00){\line(1,0){20.00}}
\put(96.33,25.00){\line(0,1){20.00}}
\put(96.33,45.00){\line(1,0){10.00}}
\put(106.33,45.00){\line(0,-1){20.00}}
\put(106.33,25.00){\line(1,0){10.00}}
\put(116.33,25.00){\line(0,1){40.00}}
\put(116.33,65.00){\line(-1,0){10.00}}
\put(106.33,65.00){\line(0,-1){10.00}}
\put(16.33,65.00){\circle*{1.89}}
\put(16.33,55.00){\circle*{1.89}}
\put(16.33,45.00){\circle*{1.89}}
\put(16.33,35.00){\circle*{1.89}}
\put(16.33,25.00){\circle*{1.89}}
\put(26.33,65.00){\circle*{1.89}}
\put(26.33,55.00){\circle*{1.89}}
\put(26.33,45.00){\circle*{1.89}}
\put(26.33,35.00){\circle*{1.89}}
\put(26.33,25.00){\circle*{1.89}}
\put(36.33,65.00){\circle*{1.89}}
\put(36.33,55.00){\circle*{1.89}}
\put(36.33,45.00){\circle*{1.89}}
\put(36.33,35.00){\circle*{1.89}}
\put(36.33,25.00){\circle*{1.89}}
\put(46.33,65.00){\circle*{1.89}}
\put(46.33,55.00){\circle*{1.89}}
\put(46.33,45.00){\circle*{1.89}}
\put(46.33,35.00){\circle*{1.89}}
\put(46.33,25.00){\circle*{1.89}}
\put(56.33,65.00){\circle*{1.89}}
\put(56.33,55.00){\circle*{1.89}}
\put(56.33,45.00){\circle*{1.89}}
\put(56.33,35.00){\circle*{1.89}}
\put(56.33,25.00){\circle*{1.89}}
\put(76.33,65.00){\circle*{1.89}}
\put(76.33,55.00){\circle*{1.89}}
\put(76.33,45.00){\circle*{1.89}}
\put(76.33,35.00){\circle*{1.89}}
\put(76.33,25.00){\circle*{1.89}}
\put(86.33,65.00){\circle*{1.89}}
\put(86.33,55.00){\circle*{1.89}}
\put(86.33,45.00){\circle*{1.89}}
\put(86.33,35.00){\circle*{1.89}}
\put(86.33,25.00){\circle*{1.89}}
\put(96.33,65.00){\circle*{1.89}}
\put(96.33,55.00){\circle*{1.89}}
\put(96.33,45.00){\circle*{1.89}}
\put(96.33,35.00){\circle*{1.89}}
\put(96.33,25.00){\circle*{1.89}}
\put(106.33,65.00){\circle*{1.89}}
\put(106.33,55.00){\circle*{1.89}}
\put(106.33,45.00){\circle*{1.89}}
\put(106.33,35.00){\circle*{1.89}}
\put(106.33,25.00){\circle*{1.89}}
\put(116.33,65.00){\circle*{1.89}}
\put(116.33,55.00){\circle*{1.89}}
\put(116.33,45.00){\circle*{1.89}}
\put(116.33,35.00){\circle*{1.89}}
\put(116.33,25.00){\circle*{1.89}}
\put(16.33,10.00){\makebox(0,0)[lb]{(a)}}
\put(76.33,10.00){\makebox(0,0)[lb]{(b)}}
\put(46.36,65.00){\line(1,0){9.97}}
\put(46.36,65.00){\line(1,0){9.97}}
\put(56.33,65.00){\line(0,0){0.00}}
\put(23.03,38.00){\makebox(0,0)[cc]{1}}
\put(23.03,47.97){\makebox(0,0)[cc]{2}}
\put(23.03,57.95){\makebox(0,0)[cc]{3}}
\put(33.01,57.95){\makebox(0,0)[cc]{4}}
\put(36.38,65.00){\line(0,-1){9.97}}
\put(26.36,54.99){\line(0,-1){19.97}}
\put(46.33,35.02){\line(0,-1){9.98}}
\put(33.04,67.95){\makebox(0,0)[cc]{5}}
\put(22.95,67.95){\makebox(0,0)[cc]{6}}
\put(12.97,67.95){\makebox(0,0)[cc]{7}}
\put(12.97,57.97){\makebox(0,0)[cc]{8}}
\put(12.97,47.98){\makebox(0,0)[cc]{9}}
\put(12.97,38.00){\makebox(0,0)[cc]{10}}
\put(12.97,27.90){\makebox(0,0)[cc]{11}}
\put(22.95,28.01){\makebox(0,0)[cc]{12}}
\put(33.15,28.01){\makebox(0,0)[cc]{13}}
\put(33.04,38.00){\makebox(0,0)[cc]{14}}
\put(33.04,47.98){\makebox(0,0)[cc]{15}}
\put(43.03,47.98){\makebox(0,0)[cc]{16}}
\put(43.03,57.97){\makebox(0,0)[cc]{17}}
\put(43.03,67.95){\makebox(0,0)[cc]{18}}
\put(53.01,67.95){\makebox(0,0)[cc]{19}}
\put(53.01,57.97){\makebox(0,0)[cc]{20}}
\put(53.01,47.98){\makebox(0,0)[cc]{21}}
\put(53.01,38.00){\makebox(0,0)[cc]{22}}
\put(53.01,28.01){\makebox(0,0)[cc]{23}}
\put(43.03,28.01){\makebox(0,0)[cc]{24}}
\put(43.03,38.00){\makebox(0,0)[cc]{25}}
\put(83.03,38.00){\makebox(0,0)[cc]{1}}
\put(83.03,47.97){\makebox(0,0)[cc]{2}}
\put(83.03,57.95){\makebox(0,0)[cc]{3}}
\put(93.01,57.95){\makebox(0,0)[cc]{4}}
\put(93.04,67.95){\makebox(0,0)[cc]{5}}
\put(82.95,67.95){\makebox(0,0)[cc]{6}}
\put(72.97,67.95){\makebox(0,0)[cc]{7}}
\put(72.97,57.97){\makebox(0,0)[cc]{8}}
\put(72.97,47.98){\makebox(0,0)[cc]{9}}
\put(72.97,38.00){\makebox(0,0)[cc]{10}}
\put(72.97,27.90){\makebox(0,0)[cc]{11}}
\put(82.95,28.01){\makebox(0,0)[cc]{12}}
\put(93.15,28.01){\makebox(0,0)[cc]{13}}
\put(93.04,38.00){\makebox(0,0)[cc]{14}}
\put(93.04,47.98){\makebox(0,0)[cc]{15}}
\put(103.03,47.98){\makebox(0,0)[cc]{16}}
\put(113.01,47.98){\makebox(0,0)[cc]{21}}
\put(102.99,38.01){\makebox(0,0)[cc]{17}}
\put(102.99,27.94){\makebox(0,0)[cc]{18}}
\put(113.05,27.94){\makebox(0,0)[cc]{19}}
\put(113.05,38.01){\makebox(0,0)[cc]{20}}
\put(113.05,58.02){\makebox(0,0)[cc]{22}}
\put(113.05,67.97){\makebox(0,0)[cc]{23}}
\put(102.99,67.97){\makebox(0,0)[cc]{24}}
\put(102.99,58.02){\makebox(0,0)[cc]{25}}
\end{picture}

%% file: fig6.tex
%TexCad Options
%\grade{\off}
%\emlines{\off}
%\beziermacro{\on}
%\reduce{\on}
%\snapping{\off}
%\quality{2.00}
%\graddiff{0.01}
%\snapasp{1}
%\zoom{1.00}
\unitlength 0.40mm
\linethickness{1.0pt}
\begin{picture}(140.00,131.93)
\put(30.00,30.00){\circle*{5.20}}
\put(70.00,30.00){\circle*{5.20}}
\put(110.00,30.00){\circle*{5.20}}
\put(30.00,69.67){\circle*{5.20}}
\put(30.00,109.33){\circle*{5.20}}
\put(70.00,69.67){\circle*{5.20}}
\put(70.00,109.33){\circle*{5.20}}
\put(110.00,69.67){\circle*{5.20}}
\put(110.00,109.33){\circle*{5.20}}
\put(40.00,40.00){\circle*{5.20}}
\put(50.00,50.00){\circle*{5.20}}
\put(80.00,40.00){\circle*{5.20}}
\put(90.00,50.00){\circle*{5.20}}
\put(120.00,40.00){\circle*{5.20}}
\put(130.00,50.00){\circle*{5.20}}
\put(40.00,79.67){\circle*{5.20}}
\put(50.00,89.67){\circle*{5.20}}
\put(40.00,119.33){\circle*{5.20}}
\put(50.00,129.33){\circle*{5.20}}
\put(80.00,79.67){\circle*{5.20}}
\put(90.00,89.67){\circle*{5.20}}
\put(80.00,119.33){\circle*{5.20}}
\put(90.00,129.33){\circle*{5.20}}
\put(120.00,79.67){\circle*{5.20}}
\put(130.00,89.67){\circle*{5.20}}
\put(120.00,119.33){\circle*{5.20}}
\put(130.00,129.33){\circle*{5.20}}
\put(30.00,30.00){\line(1,0){80.00}}
\put(110.00,30.00){\line(1,1){10.33}}
\put(120.33,40.33){\line(-1,0){80.00}}
\put(40.33,40.33){\line(1,1){9.67}}
\put(50.00,50.00){\line(1,0){80.00}}
\put(130.00,50.00){\line(0,1){40.00}}
\put(130.00,90.00){\line(-1,-1){20.00}}
\put(110.00,70.00){\line(0,1){40.00}}
\put(110.00,110.00){\line(1,1){20.00}}
\put(130.00,130.00){\line(-1,0){40.00}}
\put(90.00,130.00){\line(-1,-1){10.00}}
\put(80.00,120.00){\line(0,-1){40.00}}
\put(80.00,80.00){\line(1,1){10.00}}
\put(90.00,90.00){\line(-1,0){40.00}}
\put(50.00,90.00){\line(0,1){40.00}}
\put(50.00,130.00){\line(-1,-1){20.00}}
\put(30.00,110.00){\line(1,0){40.00}}
\put(70.00,110.00){\line(0,-1){40.00}}
\put(70.00,70.00){\line(-1,0){40.00}}
\put(30.00,70.00){\line(1,1){10.00}}
\put(140.00,130.00){\makebox(0,0)[cc]{$C$}}
\put(140.00,90.00){\makebox(0,0)[cc]{$L$}}
\put(86.67,73.67){\makebox(0,0)[cc]{$O$}}
\put(72.33,124.00){\makebox(0,0)[cc]{$F$}}
\end{picture}

%% file: fig7.tex
%TexCad Options
%\grade{\off}
%\emlines{\off}
%\beziermacro{\on}
%\reduce{\off}
%\snapping{\off}
%\quality{2.00}
%\graddiff{0.01}
%\snapasp{1}
%\zoom{1.00}
\unitlength 1.00mm
\linethickness{1.0pt}
\begin{picture}(119.02,79.97)
\put(70.00,79.97){\line(0,-1){1.95}}
\put(70.00,75.98){\line(0,-1){1.94}}
\put(70.00,72.00){\line(0,-1){1.94}}
\put(70.00,68.02){\line(0,-1){1.94}}
\put(70.00,64.04){\line(0,-1){1.95}}
\put(70.00,60.06){\line(0,-1){1.95}}
\put(70.00,56.07){\line(0,-1){1.94}}
\put(70.00,52.09){\line(0,-1){1.94}}
\put(70.00,48.11){\line(0,-1){1.94}}
\put(70.00,44.13){\line(0,-1){1.95}}
\put(70.00,40.15){\line(0,-1){1.95}}
\put(70.00,36.16){\line(0,-1){1.94}}
\put(70.00,32.18){\line(0,-1){1.94}}
\put(70.00,28.20){\line(0,-1){1.94}}
\put(40.00,70.00){\line(1,0){35.00}}
\put(40.00,40.00){\line(1,0){35.00}}
\put(80.00,60.00){\line(1,0){15.00}}
\put(84.99,50.03){\line(1,0){15.04}}
\put(82.54,55.00){\line(1,0){14.97}}
\put(82.51,24.97){\line(1,0){17.51}}
\put(79.98,30.03){\line(1,0){17.57}}
\put(44.99,68.03){\vector(0,1){0.20}}
\put(44.99,55.00){\line(0,1){13.03}}
\put(44.99,41.97){\vector(0,-1){0.20}}
\put(44.99,55.00){\line(0,-1){13.03}}
\put(42.00,55.00){\makebox(0,0)[rc]{$\Delta E$}}
\put(37.33,40.00){\makebox(0,0)[rc]{$E_0$}}
\put(105.33,25.00){\makebox(0,0)[lc]{$E_0$}}
\put(114.97,30.01){\line(1,0){3.04}}
\put(114.97,24.95){\line(1,0){3.04}}
\put(116.53,32.03){\vector(0,-1){0.20}}
\put(116.53,33.97){\line(0,-1){1.94}}
\put(116.53,23.01){\vector(0,1){0.20}}
\put(116.53,20.99){\line(0,1){2.02}}
\put(119.02,27.43){\makebox(0,0)[lc]{$\Delta E$}}
\put(55.00,10.00){\makebox(0,0)[cc]{$\gamma=0$}}
\put(85.00,10.00){\makebox(0,0)[cc]{$\gamma \neq 0$}}
\put(71.32,25.02){\line(1,0){5.02}}
\put(71.32,50.00){\line(1,0){5.02}}
\put(73.83,35.06){\vector(0,1){2.89}}
\put(73.83,30.04){\vector(0,-1){3.01}}
\put(73.83,62.55){\vector(0,1){5.40}}
\put(73.83,57.53){\vector(0,-1){5.52}}
\put(73.83,60.04){\makebox(0,0)[cc]{$\delta e_1$}}
\put(73.95,32.55){\makebox(0,0)[cc]{$\delta e_0$}}
\put(75.04,70.02){\line(1,-2){9.99}}
\put(74.98,39.99){\line(1,-2){7.51}}
\end{picture}